\documentclass[aps,prl,reprint,letterpaper,showpacs,amsmath,amssymb,longbibliography]{revtex4-1}
\usepackage{graphicx}
\usepackage[absolute]{textpos}
\begin{document}
\begin{textblock*}{8.5in}(0in,0.25in)
\begin{center}
PHYSICAL REVIEW LETTERS
\end{center}
\end{textblock*}
\begin{textblock*}{2in}(0.6in,0.4in)
PRL \textbf{114}, 104501 (2015)
\end{textblock*}
\begin{textblock*}{2in}(6.92in,0.26in)
{\footnotesize week ending\\13 MARCH 2015}
\end{textblock*}
\begin{textblock*}{7.2in}(0.6in,0.55in)
\hrulefill
\end{textblock*}
\begin{textblock*}{2.5in}(5.02in,2.94in)
{\footnotesize \doi{10.1103/PhysRevLett.114.104501}}
\end{textblock*}
\begin{textblock*}{2.5in}(5.6in,10.5in)
\copyright2015 American Physical Society
\end{textblock*}
\title{Theory of the Antibubble Collapse}
\received{3 September 2014}\published{10 March 2015}
\author{Denis Nikolaevich \surname{Sob'yanin}}
\email{sobyanin@lpi.ru}
\affiliation{I. E. Tamm Division of Theoretical Physics, P. N. Lebedev Physical Institute of the Russian Academy of Sciences,\\Leninskii Prospekt 53, Moscow 119991, Russia}
\begin{abstract}
A theory of the collapse of a punctured antibubble is developed. The motion of the rim of air formed at the edge of the collapsing air film cannot be described by a potential flow and is characterized by high Reynolds numbers. The rim velocity is not constant but gradually decreases with time and is determined by the balance between the surface tension and hydrodynamic drag forces. A collapse equation is derived and solved. The agreement between the theory and existing experiments is shown.
\end{abstract}
\maketitle

An antibubble is a thin spherical gas film containing and being surrounded by a liquid. It is a peculiar antipode to an ordinary bubble, a spherical liquid film with a gas inside and outside. The study of antibubbles has a relation to the physics of films, interfaces, foams, bubbles, and drops \cite{GennesBrochardWyartQuere2004}. The formation of antibubbles differs from the cavitation of bubbles: antibubbles are generated by gently dripping or pouring a surfactant solution onto the surface of the same solution. Antibubbles were seemingly observed in 1931 during the investigation of soap drops on a water surface \cite{HughesHughes1932}: some of the drops sank under the water surface and showed interference colors; from this observation the authors concluded that they had observed drops surrounded by a soap film. The existence of the air film was mentioned later~\cite{Riedel1938}, and the objects studied were referred to as ``inverted soap bubbles''  \cite{Skogen1956} or ``inverse bubbles'' \cite{Baird1960}. The term ``antibubble'' was coined by Pavlov-Verevkin in his Russian paper entitled ``Soap Antibubbles'' in 1966 \cite{PavlovVerevkin1966}. Interest in antibubbles was revived in 2003, when the formation and collapse of antibubbles and the development of the concomitant fluid instabilities were observed with a high-speed video camera \cite{DorboloCapsVandewalle2003}. Since then the stabilization of antibubbles \cite{KimVogel2006,Poortinga2013}, optimal conditions for the antibubble formation \cite{KimStone2008}, and antibubble lifetime distribution \cite{DorboloEtal2005,DorboloEtal2010,ScheidEtal2012,ScheidZawalaDorbolo2014} have been studied. The formation of an ``antidrop,'' an object similar to an antibubble but in which a liquid phase replaces the air film, was also observed \cite{GalvinEtal2006}. Antibubbles not only allow one to observe and study new fast microhydrodynamic phenomena but also are interesting by their potential applications \cite{PostemaEtal2007,SilpeMcGrail2013}.

Recently, a new experimental study of the collapse of an antibubble was conducted \cite{ZouEtal2013}, in which an attempt was made to find the factors that determine the velocity of the edge of the shrinking air film after puncturing the antibubble with a pin. The authors concluded that this velocity is virtually constant. However, the earlier experimental data \cite{DorboloCapsVandewalle2003} definitely show that the velocity decreases during the collapse. This surprising paradox poses a question about our understanding of the antibubble collapse and requires a proper theoretical description of the phenomenon, still absent despite the existing experiments. The purpose of this Letter is to develop the theory of the antibubble collapse. In particular, the theory solves the above paradox, reveals what actually determines the collapse velocity, and shows that the antibubble collapse differs from the rupture of liquid films.

\begin{figure}[b]
\includegraphics[width=8.4cm]{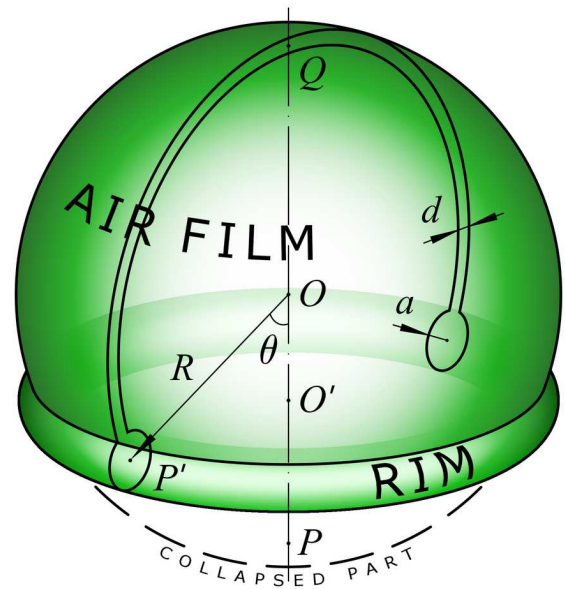}
\caption{\label{Fig1}(color online) Scheme of a collapsing antibubble.}
\end{figure}

Figure~\ref{Fig1} shows a scheme of a collapsing antibubble. The antibubble of thickness~$d$ and radius~$R$ is punctured at point~P at time $t=0$. In the experiment \cite{ZouEtal2013} the characteristic values for the antibubble thickness and radius are $d\approx 3\text{ }\mu\text{m}$ and $R\approx4$~mm, respectively; therefore, we will assume that
$d\ll R$. After puncturing, the liquids inside and outside the antibubble come into contact with each other at point~P and the air film, originally closed, starts shrinking due to the surface tension of the liquid. The film shrinking results in the appearance and expansion of a circular hole in the antibubble. The collapsing air film is axisymmetric with respect to polar axis~OP, where O is the center of the antibubble, so that the center~O$'$ of the hole always remains at OP while moving from starting point~P to diametrically opposite point~Q.

The edge of the hole represents a rim that contains all the air from the collapsed part of the antibubble. We assume, for simplicity, that the cross section of the rim by the plane passing through OP is a circle of radius~$a$; then the rim is a torus. Consideration of the rim makes sense if $a\gg d$. We also assume that $a\ll R$, which can be rewritten as $a_{\max}\ll R$, where $a_{\max}=\sqrt[3]{2dR^2/\pi}$ is the maximum rim radius that corresponds to the torus of maximum volume $2\pi^2a_{\max}^3=4\pi R^2d$.

The position of the rim is given by polar angle~$\theta$, the angle between polar axis~OP and radial vector~OP$'$, where P$'$ is a point at the guiding circle of the torus. Equating the volume of the rim, $V_\text{rim}=2\pi^2 a^2 R\sin\theta$, to the volume of the collapsed air film, $V_\text{col}=2\pi R^2d(1-\cos\theta)$, yields the dependence of the rim radius on~$\theta$:
\begin{equation}
\label{aOnTheta}
a=a_0\sqrt{\tan\frac{\theta}{2}},
\end{equation}
where $a_0=\sqrt{dR/\pi}$ is the semicollapse rim radius, the rim radius corresponding to $\theta=\pi/2$, when the first hemisphere of the antibubble has disappeared. The volume conservation condition is utilized in Refs.~\cite{DorboloCapsVandewalle2003,ZouEtal2013} to estimate the thickness of the antibubble; the possibility of using this condition is experimentally shown in Ref.~\cite{ZouEtal2013}. Equation~\eqref{aOnTheta} makes sense for $\theta$ from the range $\theta_{\min}\ll\theta\leqslant\theta_{\max}$, where $\theta_{\min}=2d/\pi R$ and $\theta_{\max}=\pi-a_{\max}/R$. The first inequality is necessary for the possibility of neglecting the difference between using $\theta$ and the more correct value $\theta+a/R$ in calculating $V_\text{col}$; it is in fact equivalent to the aforementioned inequality $a\gg d$. The angle $\theta_{\max}$ in the second inequality corresponds to the torus of maximum volume, so that $a(\theta_{\max})=a_{\max}$.

The rim moves in the liquid with velocity $v=R\,d\theta/dt$. We assume that the rim is sufficiently thin: $a\ll R\sin\theta$. If this condition is satisfied, the flow about the rim is quasi-two-dimensional. This condition implies the above condition $a\ll R$ and can be violated only at the final stage of the collapse, when the polar angle is close to~$\pi$. Thus, we should assume that $\pi-\theta\gg a/R$, or equivalently $\pi-\theta\gg\pi-\theta_{\max}$; this means that $\theta$ should not be very close to~$\theta_{\max}$.

Denote by $\rho$ the mass density and by $\eta$ the dynamic viscosity of the liquid. We can then define the corresponding Reynolds number $\text{Re}=2av\rho/\eta$. Experimentally \cite{ZouEtal2013}, $\rho\sim10^3\text{ kg}\,\,\text{m}^{-3}$, $\eta\sim10^{-3}\text{ Pa}\,\,\text{s}$, $v\sim1\text{ m}\,\text{s}^{-1}$, and $a$ can be estimated via $d$ and $R$ as $a\sim a_0\sim60\text{ }\mu\text{m}$; therefore, $\text{Re}\sim120$. The flow about the cylinder at such $\text{Re}$ is characterized by the existence of the von K\'{a}rm\'{a}n vortex street \cite{VanDyke1982} and is obviously nonpotential due to the existence of vortices.

Note in this connection that the assumption about the flow potentiality was used in Ref.~\cite{ZouEtal2013} and allowed the authors, after some order-of-magnitude estimations, to obtain a constant velocity of the rim, $v\propto\sqrt{\sigma/\rho d}$, where $\sigma$ is the surface tension. We conclude that the potential flow about the rim is not realized and that the above formula is inapplicable in the experimental conditions of Ref.~\cite{ZouEtal2013}. Particularly, $v$ need not be constant, which resolves the paradox mentioned in the introduction. Thus, the antibubble collapse is characterized by a nonpotential, vortical flow with high Reynolds numbers, and it is necessary to study which factors actually determine the velocity of the rim.

The motion of the rim is due to the surface tension. The surface tension force acting upon the unit segment of the rim is $F_\sigma=2\sigma$, with $2$ due to the existence of two interfaces between the air film and liquid. This force is counterbalanced by the hydrodynamic drag force $F_\text{d}=C_\text{d}\rho v^2 a$, where $C_\text{d}$ is the drag coefficient. For simplicity we will assume the drag coefficient constant, which implies relatively high Reynolds numbers; we may choose $C_\text{d}\approx1.1$---the typical drag coefficient for a circular cylinder at $300\leqslant\text{Re}\leqslant2\times10^5$~\cite{Michaelides2006}. We also restrict ourselves to considering only the two mentioned forces, and the comparison with experiments will show \textit{a posteriori} that other possible forces do not make a significant contribution.

The above balance between the forces can be qualitatively interpreted as thus: the surface energy of the collapsing air film turns into the kinetic energy of the liquid behind the rim. If the unit segment of the rim has moved the unit distance, the energy of the disappeared air film is $2\sigma$ whereas the volume of the liquid is~$2a$; the kinetic energy in this volume is $\sim\rho v^2 a$ because the typical energy density is $\sim\rho v^2$/2. The balance between the two energies is equivalent to the condition $F_\sigma=F_\text{d}$ with $C_\text{d}\sim1$. Such consideration makes sense only for $\text{Re}\gg1$.

From $F_\sigma=F_\text{d}$ with use of Eq.~\eqref{aOnTheta} we obtain the rim velocity as a function of~$\theta$:
\begin{equation}
\label{rimVelocity}
v=v_0\sqrt[4]{\cot\frac{\theta}{2}},
\end{equation}
where
\begin{equation}
\label{semicollapseRimVelocity}
v_0=\sqrt{\frac{2\sigma}{C_\text{d}\rho a_0}}
\end{equation}
is the semicollapse rim velocity, which corresponds to $\theta=\pi/2$. We immediately see from Eq.~\eqref{rimVelocity} that the rim velocity is not constant and decreases with time because $\theta$ obviously increases during the collapse.

Let us discuss an interesting question: Why is the rim velocity found in Ref.~\cite{ZouEtal2013} constant if the hydrodynamic drag force vanishes in a potential flow due to d'Alembert's paradox while the surface tension force is nonzero? The authors consider the balance between the rate of increasing the kinetic energy of the potential flow about the rim and the rate of decreasing the surface energy. Since the energy of potential flow about the cylinder of unit length is $m_\text{a}v^2/2$, where $m_\text{a}=\rho\pi a^2$ is the added mass~\cite{Lamb1975}, the former rate results in the action of an added mass force. This force is, however, not the usual added mass force $m_\text{a}\dot{v}$ due to acceleration, which vanishes, but the reactive force $\dot{m}_\text{a}v$ due to increasing the added mass of the rim. That is why the result of Ref.~\cite{ZouEtal2013}, $v\propto\sqrt{\sigma/\rho d}$, is similar to that of Ref.~\cite{Culick1960}: in such a consideration the air film behaves similarly to a liquid film with a massive rim.

Actually, the situation is quite different: the flow is nonpotential, and the hydrodynamic drag force is not only nonzero but also much larger than the added mass force. Both the velocity and the added mass change with time, so we estimate the latter force as $F_\text{a}=d\,m_\text{a}v/dt$. It follows from Eqs.~\eqref{aOnTheta} and \eqref{rimVelocity} that $F_\text{a}/F_\text{d}=(3\pi/4C_\text{d})(a/R\sin\theta)\ll1$. Therefore, the collapse of an air film is not similar to the collapse of a liquid film.

Let us define a typical time characterizing the collapse, $T=R/v_0$, and a dimensionless time $\tau=t/T$. From Eq.~\eqref{rimVelocity} we then get the collapse equation
\begin{equation}
\label{collapseEquation}
\frac{d\theta}{d\tau}=\sqrt[4]{\cot\frac{\theta}{2}},
\end{equation}
where $\theta$ is considered as a function of~$\tau$. Thus, Eq.~\eqref{collapseEquation} determines the time dependence of~$\theta$.

First find the semicollapse time~$t_0$, the time of the disappearance of the first hemisphere of the antibubble. Integrating $\sqrt[4]{\tan(\theta/2)}$ over $\theta$ from $0$ to $\pi/4$ yields the dimensionless semicollapse time
$\tau_0=\beta(5/8)\approx1.209$, where $\beta(x)=[\psi\boldsymbol{(}(x+1)/2\boldsymbol{)}-\psi(x/2)]/2$, $\psi(x)=d\ln\Gamma(x)/dx$ is the digamma function, and $\Gamma(x)$ is the Euler gamma function~\cite{GradshteynRyzhik2007}. The semicollapse time is then $t_0=\tau_0 T$.

Second find the collapse time~$t_\text{col}$, the time of the disappearance of the whole antibubble. Integrating $\sqrt[4]{\tan(\theta/2)}$ over $\theta$ from $0$ to $\pi/2$ yields the dimensionless collapse time $\tau_\text{col}=\pi\sec(\pi/8)\approx3.400$ \cite{GradshteynRyzhik2007}. The collapse time is then $t_\text{col}=\tau_\text{col}T$. We see again that the collapse gradually slows down: $t_\text{col}/t_0\approx2.812>2$.

\begin{figure}
\includegraphics[width=8.4cm]{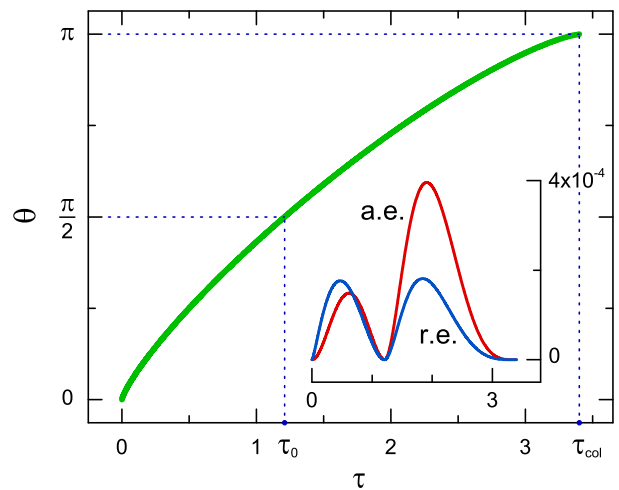}
\caption{\label{Fig2}(color online) Polar angle $\theta$ against dimensionless time $\tau$---exact solution [Eq.~\eqref{collapseSolution}] of the collapse equation [Eq.~\eqref{collapseEquation}]. Inset: absolute (a.e.) and relative (r.e.) errors of fit [Eq.~\eqref{fit}] against~$\tau$.}
\end{figure}

Now we turn our attention to the time dependence of~$\theta$. With the substitution $y(\theta)=\sqrt[4]{\tan(\theta/2)}$
we derive an analytical solution of the collapse equation [Eq.~\eqref{collapseEquation}] in the form $\tau=\tau(\theta)$:
\begin{eqnarray}
\tau(\theta)&=&\cos\frac{\pi}{8}\sum_{n=0}^1\biggl\{2\arctan\biggl[y(\theta)\csc\frac{\pi}{8}
+(-1)^n\cot\frac{\pi}{8}\biggr]
\nonumber\\
& &+(-1)^n\ln\biggl[y(\theta)^2+(-1)^n2y(\theta)\sin\frac{\pi}{8}+1\biggr]\biggr\}
\nonumber\\
& &-\sin\frac{\pi}{8}\sum_{n=0}^1\biggl\{2\arctan\biggl[y(\theta)\sec\frac{\pi}{8}
+(-1)^n\tan\frac{\pi}{8}\biggr]
\nonumber\\
& &+(-1)^n\ln\biggl[y(\theta)^2+(-1)^n2y(\theta)\cos\frac{\pi}{8}+1\biggr]\biggr\}
.\label{collapseSolution}
\end{eqnarray}
This solution is presented in Fig.~\ref{Fig2}.

The exact solution [Eq.~\eqref{collapseSolution}] is rather cumbersome, and we need a tractable analytical fit to easily work with the inverse function $\theta=\theta(\tau)$. We see directly from the collapse equation [Eq.~\eqref{collapseEquation}] that $\theta$ has asymptotics
\begin{equation}
\label{polarAngleAsymptotics}
\theta(\tau)=
\begin{cases}
2\,(5\tau/8)^{4/5},&\tau\rightarrow0,\\
\pi/2+(\tau-\tau_0),&\tau\rightarrow\tau_0,\\
\pi-2\,\bigl[3(\tau_\text{col}-\tau)/8\bigr]^{4/3},&\tau\rightarrow\tau_\text{col}.
\end{cases}
\end{equation}
We then construct a fit $\hat{\theta}(\tau)$ to the exact solution $\theta(\tau)$ from the condition that $\hat{\theta}(\tau)$ satisfies the exact asymptotics [Eq.~\eqref{polarAngleAsymptotics}]:
\begin{equation}
\label{fit}
\hat{\theta}(\tau)=
\begin{cases}
A_0\tau^{4/5}(1-A\tau^\alpha),&0\leqslant\tau\leqslant\tau_0,\\
\pi-B_0(\tau_\text{col}-\tau)^{4/3}\\
\times[1+B(\tau_\text{col}-\tau)^\beta],&\tau_0<\tau\leqslant\tau_\text{col},
\end{cases}
\end{equation}
where $A_0=2\,(5/8)^{4/5}\approx1.373$, $B_0=2\,(3/8)^{4/3}\approx0.5408$, $\alpha=(2\pi/5-\tau_0)(A_0\tau_0^{4/5}-\pi/2)^{-1}\approx1.703$, $\beta=(\tau_\text{col}-\tau_0-2\pi/3)\bigl[\pi/2-B_0(\tau_\text{col}-\tau_0)^{4/3}\bigr]^{-1}\approx3.069$, $A=(1-\pi/2A_0\tau_0^{4/5})\tau_0^{-\alpha}\approx1.260\times10^{-2}$, and $B=\bigl[\pi/2B_0(\tau_\text{col}-\tau_0)^{4/3}-1\bigr](\tau_\text{col}-\tau_0)^{-\beta}\approx1.846\times10^{-3}$. The inset in Fig.~\ref{Fig2} shows the accuracy of the fit: the absolute error $\hat{\theta}-\theta<4\times10^{-4}\approx0.02^\circ$ and the relative error $(\hat{\theta}-\theta)/\theta<0.02\%$. We see that the constructed fit [Eq.~\eqref{fit}] is of high accuracy and hence can be used instead of the exact solution [Eq.~\eqref{collapseSolution}].

\begin{figure}
\includegraphics[width=8.4cm]{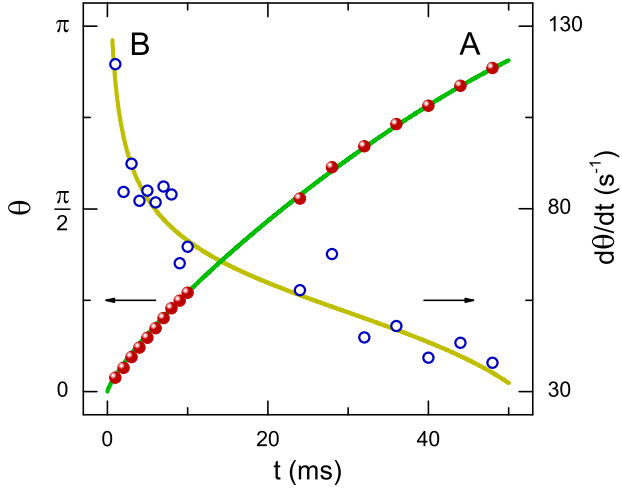}
\caption{\label{Fig3}(color online) Polar angle $\theta$ (A) and angle velocity $d\theta/dt$ (B) against time $t$---experimental data \cite{DorboloCapsVandewalle2003} (balls and circles) and theoretical curves (solid lines).}
\end{figure}

Figure~\ref{Fig3} shows the data from Ref.~\cite{DorboloCapsVandewalle2003} on the time dependence of polar angle $\theta$ and the theoretical curve $\hat{\theta}(t/T)$ (A). We observe a good agreement between the theory and experiment. Least squares fitting gives the characteristic time $T=18.06\pm0.08$~ms (standard error shown),  from which we calculate the semicollapse time $t_0=21.8\pm0.1$~ms and collapse time $t_\text{col}=61.4\pm0.3$~ms.

Figure~\ref{Fig3} also shows the data from Ref.~\cite{DorboloCapsVandewalle2003} on the time dependence of angle velocity $d\theta/dt$ and the theoretical curve $\{\cot[\hat{\theta}(t/T)/2]\}^{1/4}/T$ (B). We again observe a good agreement, and we can independently calculate $T$ from these data: $T=16.8\pm0.5$~ms. This value is consistent with the above value, but has a much higher uncertainty because the experimental error of $d\theta/dt$ is much higher than that of~$\theta$. These errors can be estimated from the respective fits: $\sigma_\theta\approx0.02$ and $\sigma_{d\theta/dt}\approx7\text{ s}^{-1}$, and correspond to the typical relative errors $\delta_\theta\approx1.3\%$ and $\delta_{d\theta/dt}\approx12\%$ (calculated at the semicollapse point, where $\theta=\pi/2$ and $d\theta/dt=T^{-1}$). Thus, due to smaller experimental errors, the data on the time dependence of $\theta$ contain more precise information about various temporal characteristics of the collapse.

\begin{figure}
\includegraphics[width=8.4cm]{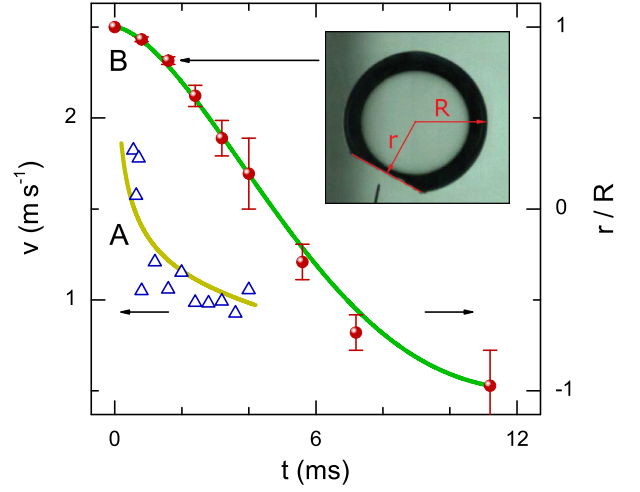}
\caption{\label{Fig4}(color online) Rim velocity $v$ (A) and distance between antibubble and hole centers, $r$, over radius $R$ (B) against time $t$---experimental data \cite{ZouEtal2013} (triangles and balls) and theoretical curves (solid lines). Inset: $r$ and $R$ on the image \cite{ZouEtal2013} for the third data point.}
\end{figure}

Figure~\ref{Fig4} shows the data from Ref.~\cite{ZouEtal2013} on the time dependence of rim velocity $v$ and the theoretical curve $v_0\{\cot[\hat{\theta}(t/T)/2]\}^{1/4}$, where $v_0=R/T$ and $R=3.80$~mm (A). From least squares fitting we have $T=4.1\pm0.2$~ms and the semicollapse rim velocity $v_0=0.93\pm0.05\text{ m}\,\text{s}^{-1}$, and we estimate the absolute and relative experimental errors as $\sigma_v\approx0.2\text{ m}\,\text{s}^{-1}$ and $\delta_v\approx21\%$. These data are consistent with the fact that the rim velocity gradually decreases with time and do not require that it is constant.

Reference \cite{ZouEtal2013} also contains a chronological sequence of images of the collapsing antibubble [Fig.~2 therein]. I process these images in a manner shown in the inset in Fig.~\ref{Fig4} to obtain the data on the time dependence of $r/R$, where $r$ is the distance between the center O of the antibubble and the center O$'$ of the circular hole in the air film (Fig.~\ref{Fig1}). Note that the rim is not clearly visible in the images because its small radius is much less than the antibubble radius. These data, with error bars from irregularity of the air film edge, and the theoretical curve $\cos\hat{\theta}(t/T)$ are presented in Fig.~\ref{Fig4}, curve (B), and again demonstrate a good agreement between the theory and experiment. Weighted least squares fitting gives more precise values than above: $T=3.9\pm0.1$~ms and $v_0=0.97\pm0.03\text{ m}\,\text{s}^{-1}$.

Very significantly, we can independently find the semicollapse rim velocity $v_0$ directly from Eq.~\eqref{semicollapseRimVelocity}: using the experimental values $d=2.95\text{ }\mu\text{m}$, $\rho=992\text{ kg}\,\text{m}^{-3}$, and $\sigma=0.033\text{ N}\,\text{m}^{-1}$ \cite{ZouEtal2013} and putting $C_\text{d}=1.1$, we obtain $v_0=1.01\pm0.05\text{ m}\,\text{s}^{-1}$, which is consistent with the two above values. This consistency confirms that the temporal development of the collapse observed in the experiment is correctly described by the theory, without fitting but from the known physical parameters.

When choosing $C_\text{d}$, we have considered the drag force acting upon a solid cylinder, with no-slip boundary conditions. Meanwhile, the rim is made of air, and free-slip boundary conditions may be more suitable, but result in remarkably decreasing $C_\text{d}$~\cite{LegendreLaugaMagnaudet2009}. On the other hand, we have the air rim not in pure water but in a soap solution, and surfactant molecules adsorb at the interface. Because of the film of surfactant molecules, an ordinary bubble experimentally behaves at $\text{Re}\ll1$ not as an air ball with free-slip conditions but as a solid ball with no-slip conditions~\cite{FrumkinLevich1947,Levich1962}, and a similar effect may be expected in the case of the rim. However, the effect can be reduced because the boundary of the moving rim goes into the two fixed interfaces of the air film before the rim. Thus, it is reasonable to consider $C_\text{d}$ as an adjustable parameter.

We have seen that $C_\text{d}\sim1$ agrees with the experimental data and hence could argue in favor of no-slip conditions. Interestingly, the theory allows us to calculate $C_\text{d}$ from experiment with the help of Eq.~\eqref{semicollapseRimVelocity}. I propose to study experimentally how $C_\text{d}$ depends on various physical parameters of the antibubble and on the surfactant concentration, which can shed some light on the problem of boundary conditions.

In conclusion, I have developed the theory of the collapse of an antibubble. After the antibubble is punctured, the air film  starts shrinking due to the surface tension of the liquid, which results in the appearance and expansion of a circular hole in the antibubble. The collapse is characterized by a nonpotential flow about the rim that forms at the edge of the shrinking air film, and the flow has high Reynolds numbers. The rim velocity does not vary linearly with $\sqrt{\sigma/\rho d}$ and is not constant; therefore, the collapse of an antibubble differs from the rupture of a liquid film \cite{Culick1960}. The velocity gradually decreases with time and is determined by the balance between the surface tension and hydrodynamic drag forces acting upon the rim. The collapse occurs so that the surface energy of the collapsing air film turns into the kinetic energy of the liquid behind the rim. I have derived and solved the collapse equation, the solution of which gives the time dependence of the polar angle and describes the temporal development of the collapse. I have demonstrated the agreement between the theory and currently available experimental data. In regard to an interesting problem of single bubble sonoluminescence \cite{MagnaudetLegendre1998,SadighiBonabiRezaeiNasirabadGalavani2009,SadighiBonabiEtal2011}, I propose to study how the antibubble collapse occurs in the presence of an acoustic field.

\begin{acknowledgments}
I would like to thank St\'{e}phane Dorbolo for providing me with the experimental data presented in Fig.~4 of Ref.~\cite{DorboloCapsVandewalle2003}. I would also like to thank Jens Eggers for useful advice.
\end{acknowledgments}
\providecommand{\noopsort}[1]{}\providecommand{\singleletter}[1]{#1}%
\end{document}